\documentclass[draft]{agujournal2019}
\usepackage{url} 
\usepackage{lineno}

\usepackage[inline]{trackchanges} 
\usepackage{soul}
\usepackage{enumerate}

\draftfalse

\journalname{Journal of Geophysical Research: Planets}

\begin{document}


\title{Jupiter's UV auroral response to a magnetospheric compression event}


\authors{R. S. Giles\affil{1}, T. K. Greathouse\affil{1}, R. W. Ebert\affil{1,2}, W. S. Kurth\affil{3}, C. K. Louis\affil{4}, M. F. Vogt\affil{5}, B. Bonfond\affil{6}, D. Grodent\affil{6}, J.-C. Gérard\affil{6}, G. R. Gladstone\affil{1}, J. A. Kammer\affil{1}, V. Hue\affil{7}, R. J. Wilson\affil{8}, S. J. Bolton\affil{1}, J. E. P. Connerney\affil{9}}

\affiliation{1}{Space Science and Engineering Division, Southwest Research Institute, San Antonio, Texas, USA}
\affiliation{2}{Department of Physics and Astronomy, University of Texas at San Antonio, San Antonio, Texas, USA}
\affiliation{3}{Department of Physics and Astronomy, University of Iowa, Iowa City, Iowa, USA}
\affiliation{4}{LESIA, Observatoire de Paris, Université PSL, CNRS, Sorbonne Université, Université de Paris, Meudon, France}
\affiliation{5}{Planetary Science Institute, USA}
\affiliation{6}{Laboratoire de Physique Atmosphérique et Planétaire, STAR Institute, Université de Liège, Liège, Belgium}
\affiliation{7}{CNRS, CNES, Institut Origines, LAM, Aix-Marseille Université, Marseille, France}
\affiliation{8}{Laboratory of Atmospheric and Space Physics, University of Colorado Boulder, Boulder, Colorado, USA}
\affiliation{9}{NASA Goddard Spaceflight Center, Greenbelt, Maryland, USA}

\correspondingauthor{R. S. Giles}{rohini.giles@swri.org}


\begin{keypoints}
\item The Juno mission allows us to simultaneously compare the compression state of the magnetosphere with the total UV auroral power.
\item On December 6--7 2022, Jupiter experienced an extreme magnetospheric compression and a significant auroral brightening to 12 TW.
\item On this occasion, the auroral brightening was likely caused by a solar wind shock compressing the magnetosphere.
\end{keypoints}


\begin{abstract}
The highly elliptical polar orbit of the Juno mission provides a unique opportunity to simultaneously measure the compression state of Jupiter's magnetosphere and the total power emitted by the planet's ultraviolet aurora, using a single spacecraft. This allows us to study how Jupiter's aurora respond to a compression event. In this paper, we present a case study of an extreme compression event that occurred on December 6--7 2022 when Juno was a distance of 70 R\textsubscript{J} from Jupiter. This extreme compression was accompanied by a very large increase in the ultraviolet auroral emissions to 12 TW, a factor of six higher than the baseline level. This event coincided with the predicted arrival of a powerful interplanetary shock, which was expected to cause the largest increase in the solar wind dynamic pressure seen thus far during the Juno mission. The simultaneous occurrence of the interplanetary shock, the extreme compression and the bright ultraviolet aurora suggests that in this case, the auroral brightening was caused by the solar wind shock compressing the magnetosphere.
\end{abstract}


\section*{Plain Language Summary}

NASA's Juno mission is able to both measure the power emitted by Jupiter’s ultraviolet auroras and to use magnetopause crossings to measure the instantaneous size of Jupiter's magnetosphere. On December 6--7 2022, Jupiter's magnetosphere underwent an extreme compression event and this was accompanied by a large increase in the ultraviolet auroral emissions. This event coincided with a sudden disturbance in the solar wind at Jupiter. The auroral brightening was likely caused by a solar wind shock compressing the magnetosphere. 


\section{Introduction}

The total power emitted by Jupiter's ultraviolet (UV) auroras can vary significantly with time. UV observations obtained from the International Ultraviolet Explorer~\cite{prange01}, the Hubble Space Telescope~\cite<e.g.>{grodent03,clarke09,bonfond12,nichols17,yao22}, the Hisaki space observatory~\cite{kimura15,kita16}, the Cassini mission~\cite{gurnett02,pryor05} and the Juno mission~\cite{gladstone17b} have all shown intermittent enhancements in the total H\textsubscript{2} emitted power. These brightening events can last for hours to days and can reach powers that are a factor of 3--4 above the baseline level. The driving factors behind these increased emissions are still not fully understood, and there are both external and internal mechanisms that could be responsible. The solar wind conditions upstream of Jupiter could act as an external driver, with an increase or decrease in the solar wind dynamic pressure causing a compression or expansion of the magnetosphere~\cite{cowley07,feng22}. Internally, auroral variability could be caused by mass loading and unloading in the middle and outer magnetosphere, driven by plasma originating from the Io torus~\cite{bonfond12,bonfond21}. 

Contemporaneous measurements of the total auroral emission, the solar wind conditions at Jupiter, and the size of the magnetosphere are critical to understanding an external mechanism. However, these three measurements are all challenging to obtain. UV observations obtained from Earth-orbiting satellites or spacecraft in the equatorial plane of the planet are limited by the observing geometry; the entire aurora cannot be observed continuously, so a time series of the auroral brightness has a periodicity caused by the rotation of the planet. Although this periodicity can be corrected for~\cite<e.g>{clarke09,nichols17}, this observing geometry adds uncertainty to measurements of temporal variability, particularly on the scale of $\sim$0.5 Jupiter rotations. The solar wind conditions at Jupiter can only be directly measured by a spacecraft that is close to Jupiter but outside of its magnetosheath, which is a relatively rare occurrence; the solar wind can be extrapolated from measurements made at Earth using a magnetohydrodynamic (MHD) model, but there are considerable associated uncertainties~\cite{zieger08,clarke09}. Similarly, the size of the magnetosphere can only be measured by a spacecraft crossing the magnetopause boundary, and even that is only a single-point measurement; assumptions must be made in order to estimate the global extent of the magnetosphere~\cite{joy02,ranquist20}. 

There have been several studies that have compared UV auroral observations with solar wind conditions derived from MHD models. \citeA{clarke09} used UV observations from the Hubble Space Telescope (HST) to correlate the auroral power emitted by the auroras of both Jupiter and Saturn with the MHD-modeled solar wind pressure; they found that while there was a strong positive correlation at Saturn, the relationship at Jupiter was weaker, indicating that while the solar wind may have some influence on the aurora, there also appears to be activity that is independent of the solar wind. Observations made by the far-ultraviolet spectrometer EXCEED (Extreme Ultraviolet Spectroscope for Exospheric Dynamics) on the Hisaki satellite have been used to classify two different types of auroral brightening events with different origins.~\citeA{kimura15} presented short-lived (\textless10 hour) brightening events that reoccur every few days and do not appear to show any correlation with modeled solar activity, while~\citeA{kita16} described longer lasting (2--3 day) events that do correlate with solar wind shock arrival. They suggest that the former may be driven by internal processes, while the latter are externally-forced.

Spacecraft in the vicinity of Jupiter have provided valuable opportunities to study the auroral response to solar wind activity. In late 2000, the Cassini spacecraft passed by Jupiter at the same time that the Galileo spacecraft was in orbit around the planet, allowing simultaneous observations to be made from two different locations in Jupiter's vicinity. \citeA{gurnett02} used in situ solar wind measurements from both spacecraft in order to detect the arrival of several interplanetary shocks and measure their propagation speed. \citeA{gurnett02} and \citeA{pryor05} found that the arrival of the shocks at Jupiter coincided with an increase in both the hectometric radio emissions and the extreme ultraviolet auroral emissions, with the peaks occurring at the same time as maximum solar wind density. 

Most recently, the Juno spacecraft has provided opportunities to make in situ measurements of the solar wind conditions at Jupiter. In 2016, the Juno spacecraft was approaching Jupiter but was still outside of the planet's magnetosheath. This allowed the in situ plasma instrument, Jovian Auroral Distributions Experiment (JADE), to directly measure the solar wind ram pressure~\cite{wilson18}, which was then compared to the Jovian auroral emissions by \citeA{gladstone17b} and \citeA{nichols17}. \citeA{nichols17} monitored the UV aurora using HST observations and found that compression events led to increased main auroral emissions for 1--3 days afterwards. \citeA{gladstone17b} used observations made by the Juno Ultraviolet Spectrograph (UVS) and found that out of four discrete brightening events observed, only one of these events was correlated with a transient increase in the solar ram pressure, suggesting that some brightening events are not caused by external forcing. 

Measurements from the Juno spacecraft were also recently used by~\citeA{yao22} to correlate auroral activity with magnetospheric compression conditions. During the early part of the Juno mission (2016--2017), the highly elliptical orbit of the spacecraft was oriented such that apojove was close to the nominal magnetopause location on the dawn side, which meant that the spacecraft occasionally crossed the magnetopause when the magnetosphere was in a compressed state. The JADE, Waves and MAG instruments on the spacecraft are all able to detect these magnetopause crossings, and therefore measure the instantaneous size of the magnetosphere.~\citeA{yao22} compared these magnetopause crossing measurements with UV observations of the aurora from HST and determined that main auroral brightenings systematically appear under a compressed magnetopause, while auroral dawn storms, another form of enhanced auroral emission, are not correlated with magnetospheric compression. 

In this paper, we take advantage of the orbital geometry of Juno in the extended mission in order to study the response of Jupiter's UV auroras to a powerful compression event. Unlike all previous observing campaigns that have monitored Jupiter's UV auroral brightness, synoptic observations obtained by Juno UVS since 2022 directly measure the total emission from the southern aurora as a function of time without any modulation due to the planet's rotation. Combined with the contemporaneous magnetopause crossing data from Juno's in situ instruments, this provides a unique dataset that can be used to understand the relationship between the UV auroral brightness and the state of compression of the magnetosphere. In particular, this allows us to examine the timing of the auroral response, and whether the auroral brightening occurs as the magnetosphere is compressing, or while it is expanding. On December 6--7 2022, Jupiter's magnetosphere experienced an extremely powerful compression event. The MHD-modeled solar wind dynamic pressure at Jupiter reached the highest level seen thus far during the Juno mission, and this fortunately coincided with both a period where the Juno spacecraft was well-placed in its orbit to measure the location of the magnetopause, and a period of time when UVS was obtaining a synoptic observation. In this paper, we present a case study of Jupiter's magnetospheric and auroral response to this compression event. 
 

\section{Data}
\label{sec:data}

Juno is a spin-stabilized spacecraft that has been in a highly-elliptical polar orbit around Jupiter since July 2016~\cite{bolton17}. In December 2022, the orbital period was 38 days, with the perijove occurring approximately 4000 km above the planet's cloud tops and the apojove occurring at a distance of 90 R\textsubscript{J}.  In this study, we make use of three instruments on the mission: the Ultraviolet Spectrograph~\cite<UVS,>{gladstone17}, the Jovian Auroral Distributions Experiment~\cite<JADE,>{mccomas17} and the Waves investigation~\cite{kurth17}.  

\begin{figure}
\centering
\includegraphics[width=14cm]{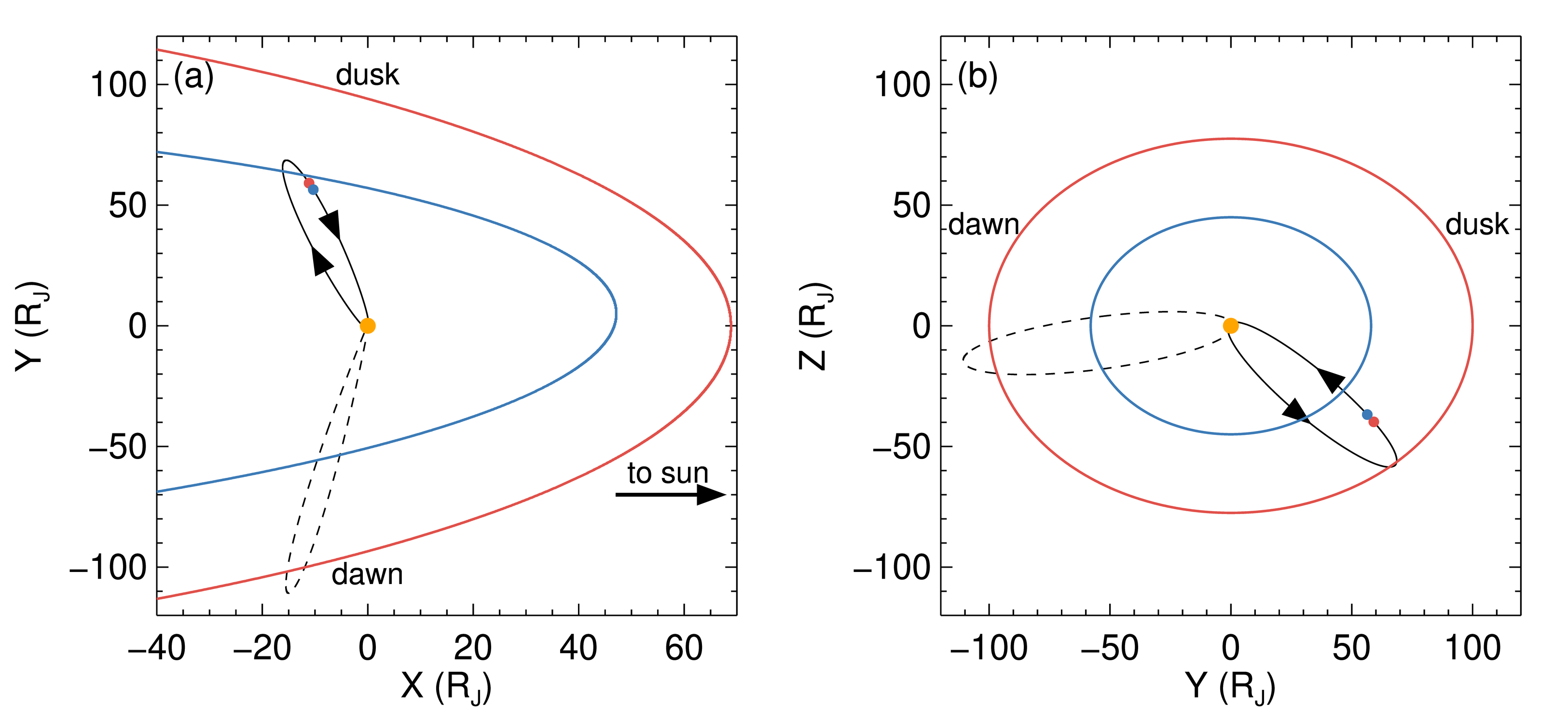}
\caption{The orbital geometry of the Juno spacecraft during the compression event in December 2022, in Jupiter-Sun-Orbit (JSO) co-ordinates. (a) is projected onto the $z=0$ plane and (b) is projected onto the $x=0$ plane. The spacecraft trajectory in late 2022 is shown by the solid black lines and the spacecraft trajectory in late 2016 is shown by the dashed black lines. Jupiter is represented by the yellow point and the red and blue points on the trajectory mark the position of the spacecraft at the time of the first and second magnetopause crossings discussed in this paper. The red and blue lines illustrate the nominal shape of the magnetopause in two different compression states (P\textsubscript{D}=0.2 nPa and P\textsubscript{D}=1.1 nPa), using the model from~\citeA{joy02} with the assumed polar flattening from~\citeA{ranquist20}.}
\label{fig:juno_trajectory}
\end{figure}

Figure~\ref{fig:juno_trajectory} shows Juno's orbit trajectory; the solid black lines show the trajectory in late 2022 (between perijoves 46 and 47), and the dashed black lines show the trajectory in late 2016 (between perijoves 2 and 3) for comparison. Figure~\ref{fig:juno_trajectory}(a) shows the trajectory projected onto the equatorial plane. At the beginning of the mission, the orbit perijove was on the dusk side of the planet, and the orbit apojove was on the dawn side. Due to the shape of the magnetosphere, shown in red and blue in two different states of compression, the spacecraft crossed the magnetopause frequently between July 2016 and August 2018~\cite{hospodarsky17,ranquist19}. As the orbit precessed, the apojove moved to the nightside of the planet; this downtail orbit precluded any magnetopause crossings, even in a compressed state, and the position of the sun relative to Jupiter prevented UVS synoptic observations. The apojove is now on the dusk side of the planet, leading to increasingly frequent magnetopause crossings from December 2021 onwards~\cite{louis23}. This once again allows us to make direct measurements of the size of Jupiter's magnetosphere and any compression events, and the solar orientation also allows for periodic UVS synoptic observations throughout the orbit.

Figure~\ref{fig:juno_trajectory}(b) shows the trajectory projected onto the $x=0$ plane. In addition to moving from dawn to dusk, the orbit has become increasingly tilted with respect to the equatorial plane; at the beginning of the mission, both perijove and apojove were close to the equator, but in December 2022, the sub-spacecraft Jovian latitude at apojove was 35$^{\circ}$S. This southern orientation has the advantage that UVS is able to view the entire southern aurora during its synoptic observation, allowing it to track the total auroral power emitted. This combination of a dawn-dusk orbit and a southern apojove means that for the first time in the mission, Juno's in situ instruments can measure the size and conditions of the outer magnetosphere while remote sensing instruments can measure the state of the entire aurora.

The red and blue points in Figure~\ref{fig:juno_trajectory} mark the location of the Juno spacecraft when the JADE and Waves instruments recorded magnetopause crossings on December 6--7 2022. The first crossing, when the spacecraft left the magnetosphere on December 6, is shown by the red point and the second crossing, when the spacecraft re-entered the magnetosphere 24 hours later, is shown by the blue point. Apojove (the furthest point from Jupiter in the orbit) occurred on November 26 2022, so on December 6--7, the spacecraft was moving towards Jupiter, and it was overtaken by the inward movement of the magnetopause as the magnetosphere rapidly compressed. 

Close to apojove, several back-and-forth magnetopause crossings were recorded by JADE and Waves. The red lines in Figure~\ref{fig:juno_trajectory} provide an illustration of the possible shape of the magnetopause boundary at that time, based on the $P_D$=0.2 nPa model from~\citeA{joy02}, with the addition of the polar flattening from~\citeA{ranquist20}. The blue lines similarly illustrate the possible shape of the magnetopause boundary during the compression event, based on a $P_D$=1.1 nPa with additional polar flattening. 

\begin{figure}
\centering
\includegraphics[width=14cm]{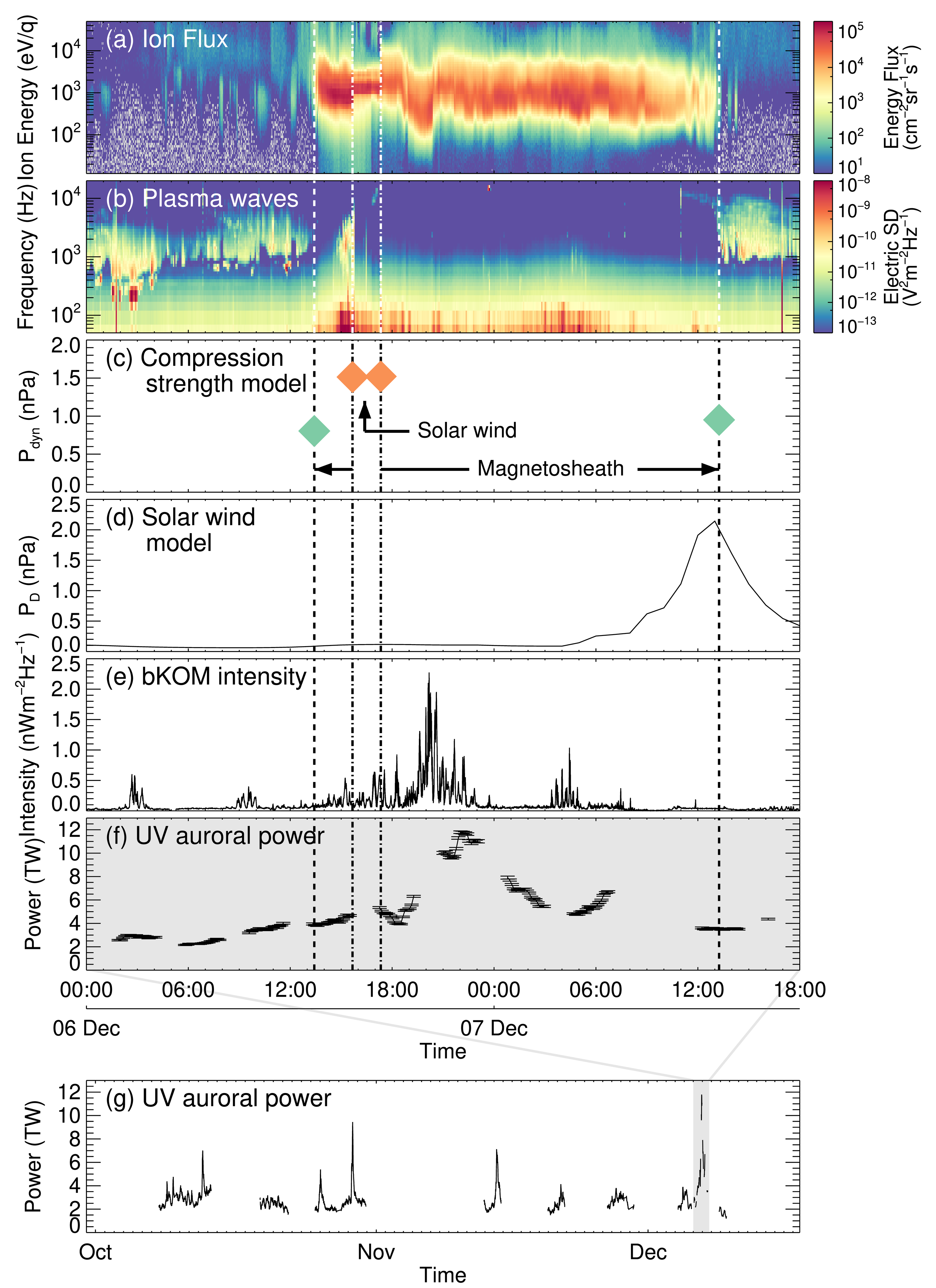}
\caption{In situ and remote sensing observations during the compression event on December 6--7. (a) Ion differential energy flux measured by JADE. (b) Electric wave spectral density measured by the Waves experiment. (c) The inferred solar wind dynamic pressure, based on the location of the spacecraft at each magnetopause and bow shock crossing~\cite{joy02}; this is a proxy for the strength of the compression. (d) The solar wind dynamic pressure modelled using the SWMF-OH MHD model. (e) bKOM radio emission measured by the Waves experiment. (f)--(g) Total UV southern aurora power measured by UVS. The vertical dashed lines and dash-dot lines in (a)--(f) show the timings of the magnetopause and bow shock crossings respectively. The grey shaded regions in (g) highlights the time period covered in (a)--(f).}
\label{fig:all_instruments}
\end{figure}

\subsection{JADE observations}

JADE provides in situ plasma ion and electron observations as the Juno spacecraft travels through Jupiter's magnetosphere. Magnetopause crossings can be detected using data from the ion sensor component of the instrument, JADE-I; as the spacecraft crosses the magnetopause from the low-density outer magnetosphere to the high-density magnetosheath, there is an abrupt and significant increase in the ion flux. This can be seen in Figure~\ref{fig:all_instruments}(a), which shows the ion differential energy flux between 00:00 on December 6 2022 and 18:00 on December 7 2022. The peak differential energy flux is less than 10\textsuperscript{2} cm\textsuperscript{-2}sr\textsuperscript{-1}s\textsuperscript{-1} until 13:25 on December 6, at which point there is a sudden increase to over 10\textsuperscript{4} cm\textsuperscript{-2}sr\textsuperscript{-1}s\textsuperscript{-1}, signifying that the spacecraft has left the magnetosphere. The ion flux remains high until almost 24 hours later, at 13:15 on December 7, when the spacecraft crosses the magnetopause again and re-enters the magnetosphere. The magnetopause crossing times are shown in Figure~\ref{fig:all_instruments} by the vertical dashed lines.

In addition to measuring the magnetopause crossing, JADE also detected a bow shock crossing, indicating that the spacecraft briefly exited the magnetosheath into the solar wind. Between 15:40 and 17:20 on December 6, the ion spectrum shown in Figure~\ref{fig:all_instruments}(a) narrows, indicating that the instrument is measuring the colder solar wind. The start and end of this time period are marked by the vertical dash-dot lines in Figure~\ref{fig:all_instruments}. During this time period, the proton velocities and densities measured by JADE suggest a solar wind dynamic pressure of approximately 0.4--0.7 nPa.

\subsection{Waves observations}
\label{sec:waves}

The Waves instrument measures the electric and magnetic components of plasma and radio waves in the environment around Jupiter. The electric field from 50 Hz to 40 MHz is measured by the single-axis electric dipole antenna and the magnetic field from 50 Hz to 20 kHz is measured by the search coil magnetometer. Figure~\ref{fig:all_instruments}(b) shows the electric component of waves in the 50 Hz to 20 kHz frequency range, measured by the Waves instrument during the same time interval as Figure~\ref{fig:all_instruments}(a). From the beginning of this time interval until 13:25 on December 6, Waves measured trapped continuum radiation with frequencies of several hundred Hz to $\sim$3 kHz. This continuum radiation is present throughout the outer magnetosphere; it is produced at both the outer edge of the Io torus and the inner edge of the magnetopause, and is trapped within the magnetospheric cavity by the relatively high density magnetosheath~\cite{scarf79,gurnett80}. At 13:25 on December 6, this radiation disappears, indicating that Juno has crossed the magnetopause into the magnetosheath, in agreement with the observations from JADE. The radiation reappears at 13:15 on December 7, indicating that Juno has re-entered the magnetosphere. 

Like JADE, Waves also detected bow shock crossings at 15:40 and 17:20 on December 6. Between these times, Waves measured electron plasma oscillations in the solar wind between 15 and 30 kHz, implying a solar wind density of 3--10 cm\textsuperscript{-3}.

Waves data can also be used to measure the radio emission from Jupiter's auroras~\cite{louis21}. Figure~\ref{fig:all_instruments}(e) shows the integrated time series for the broadband kilometric (bKOM) component of the radio emissions, which is the auroral emission between 20 and 140 kHz, and can be identified by removing the patchy narrowband kilometric (nKOM) radiation that also appears at the same frequencies but has non-auroral sources. Figure~\ref{fig:all_instruments}(e) shows that there is a significant increase in the bKOM emissions that coincides with the magnetospheric compression. 

\subsection{UVS observations}

UVS is a far-ultraviolet imaging spectrograph, covering the 68-210 nm spectral range. This spectral range was designed to target the H and H\textsubscript{2} emission bands that dominate Jupiter's auroral emissions. The instrument's 7.2$^{\circ}$-long slit is nominally oriented parallel to the spacecraft's spin axis, allowing the instrument to observe a swath of the planet during each 30-s spacecraft rotation. During the hours immediately before and after each perijove, UVS is able to combine many such swaths to build up high spatial resolution images of the aurora. In this paper, we focus on the synoptic observations that are conducted while the spacecraft is farther away from Jupiter; during these observations, the planet's angular diameter is 1--2$^{\circ}$, which is small enough to fit within the lower wide segment of the instrument's dogbone-shaped slit (see Figure S1 in the Supporting Information). 

Since October 2022, all synoptic observations have been conducted when the spacecraft has been sufficiently far south that the southern aurora can be observed continuously as the planet rotates and there is negligible periodicity due to the planet's rotation (see Figure S2 in the Supporting Information). This allows the mapped data to be converted into a time series of the total power emitted by the southern aurora. The total H\textsubscript{2} emitted power is obtained by integrating the auroral emissions observed at 155-162 nm, a spectral window with minimal atmospheric absorption, and multiplying by 8.1 \cite{bonfond17}.

Figure~\ref{fig:all_instruments}(f) shows this emitted power time series between 00:00 on December 6 2022 and 18:00 on December 7 2022, the same time interval shown in Figures~\ref{fig:all_instruments}(a)--(e). Figure~\ref{fig:all_instruments}(g) places Figure~\ref{fig:all_instruments}(f) in context, by presenting all of the synoptic UVS observations obtained between October 2022 and December 2022. 


\section{Results}

\subsection{Size of magnetosphere and magnetosheath}
\label{sec:size}

On November 26 2022, the Juno spacecraft reached its orbital apojove, at a distance of 90 R\textsubscript{J} from the planet. As noted in Section~\ref{sec:data}, several back-and-forth magnetopause crossings were recorded around this time by the JADE and Waves instruments. As Juno was returning towards Jupiter after apojove, the spacecraft initially remained within the magnetosphere. However, at 13:25 on December 6, when the spacecraft was 72 R\textsubscript{J} away from Jupiter and traveling at a speed of 3.2 km/s, it was ``overtaken'' by the magnetopause as the magnetosphere was compressed into a smaller size, illustrated by the blue lines in Figure~\ref{fig:juno_trajectory}. As noted in Section~\ref{sec:waves}, this initial magnetopause crossing was then followed by a bow shock crossing at 15:40, as the bow shock also ``overtook" the spacecraft. This marked the first time that the spacecraft had entered the solar wind since the early part of the mission when it was on the dawn flank of the magnetosphere. The spacecraft then crossed back into the magnetosheath $\sim$100 minutes later at 17:20, which could have either been due to the magnetosheath and magnetosphere beginning to expand again, or simply the rate of compression decreasing such that the spacecraft was able to ``overtake'' the bow shock. Juno then crossed back into the magnetosphere itself on December 7 at 13:15, at a distance of 68 R\textsubscript{J} from Jupiter. 

Although the Juno spacecraft spent $\sim$24 hours outside of the magnetosphere during this period of time, the compression event must have lasted even longer than this; the inward movement had already begun by the time the magnetopause passed Juno on December 6 and the outward movement continued beyond the second crossing on December 7 in order for the magnetopause to return to its original, uncompressed state. At the location of the first magnetopause crossing on December 6, the inferred $P_D$ calculated using the~\citeA{joy02} model was 0.80 nPa and at the location of the second crossing on December 7, it was 0.95 nPa. Using the bow shock model from~\citeA{joy02}, the inferred $P_D$ at the locations of the bow shock crossings were 1.51 nPa and 1.52 nPa. These inferred solar wind dynamic pressures, calculated using magnetopause and bow shock models from \citeA{joy02} are shown in Figure~\ref{fig:all_instruments}(c), with the magnetopause crossings shown in green and the bow shock crossings shown in orange. All four of these values exceed the 90th percentile values of $P_D$ provided in \citeA{joy02}.

In reality, these inferred solar wind dynamic pressures may be overestimates. The data available in \citeA{joy02} were primarily close to the equatorial plane, giving limited observational constraints on the shape of the magnetopause in the full three dimensions. Their model therefore did not include the level of polar flattening shown in~\citeA{ranquist20}. Because the magnetopause crossings presented here are farther south, the inferred $P_D$ are likely an overestimate; even in a non-compressed state, we would expect the magnetopause crossings to occur closer to Jupiter in the south than in the equatorial plane. Nevertheless, the inferred $P_D$ shown in Figure~\ref{fig:all_instruments}(c) still have qualitative value here; they act as a proxy for the extent to which the magnetosphere is compressed at a specific moment in time. The maximum recorded state of compression was at the second bow shock crossing; the true peak compression at the location of the Juno spacecraft likely occurred either between the two bow shock crossings or shortly after the second crossing, at approximately 16:00--18:00 on December 6. Even if the absolute $P_D$ values are somewhat overestimated, the simple fact that Juno was inside the magnetosphere at a distance of 90 R\textsubscript{J}, and then was later outside the magnetosphere at a distance of 68 R\textsubscript{J} shows that this was a very strong compression event. 

\subsection{Solar wind conditions}

Figure~\ref{fig:all_instruments}(d) shows the predicted solar wind dynamic pressure at Jupiter from the SWMF-OH model~\cite{zieger15}. This model was adapted from the Outer Heliosphere (OH) component of the Space Weather Modeling Framework~\cite<SWMF, >{toth12}. SWMF-OH is a two-dimensional multi-fluid MHD model that propagates solar wind data obtained at the Earth's orbit to the outer solar system. 

Figure~\ref{fig:all_instruments}(d) shows that MHD-modeling predicted a dramatic increase in the solar wind dynamic pressure at Jupiter on December 7. This suggests that the compression of Jupiter's magnetosphere was caused by the arrival of a powerful interplanetary shock. In addition to the sudden increase in the dynamic pressure, the SWMF-OH model also shows an increase in the temperature, density and velocity, suggesting that this was a Fast Forward Shock~\cite{kilpua15}. The maximum value reached 2.1 nPa and the full width at half maximum of the high-pressure pulse was approximately 4 hours. The peak velocity was 470 km/s and the peak density was 5.8 cm\textsuperscript{-3}, which is consistent with the 3--10 cm\textsuperscript{-3} measured by Waves. This dynamic pressure was the highest predicted by SWMF-OH since the arrival of Juno to Jupiter in 2016, and was a factor of 1.7 times higher than the second highest predicted P\textsubscript{D} in September 2020. ~\citeA{jackman11} found that the solar wind dynamic pressure distribution at Jupiter peaks at 0.04 nPa, with a range between 0.002 and $\sim$1 nPa, making a predicted dynamic pressure of 2.1 nPa extremely high and very rare. This pressure estimated from the SWMF-OH is higher than the 0.4--0.7 nPa estimated from the JADE data, but we note that (i) the brief time period that Juno was in the solar wind did not necessarily coincide with the peak dynamic pressure, and (ii) even a value of 0.4--0.7 nPa is an order of magnitude higher than the typical solar wind dynamic pressure. 

The precise timing of the predicted shock arrival does not correlate perfectly with the onset of the magnetospheric compression, but a timing uncertainty on the order of 24 hours is within expectations given the challenges associated with propagating solar wind conditions as far as Jupiter's orbit. The peak predicted P\textsubscript{D} occurred at 13:00 on December 7, when the magnetosphere was beginning to expand again. In order for the interplanetary shock to have caused the magnetospheric compression, it would have needed to have arrived 18---30 hours earlier, but this uncertainty in timing is consistent with statistical validation of MHD propagation models~\cite{zieger08}. 

\subsection{Auroral response}

Figure~\ref{fig:all_instruments}(g) shows the total H\textsubscript{2} emissions from Jupiter's southern aurora as a function of time. The baseline power level is approximately 2 TW, in agreement with the Juno approach observations made by~\citeA{gladstone17b}. There are several brightening events that occur during the three-month period shown in Figure~\ref{fig:all_instruments}(g), and the brightest of these coincides with the compression event on December 6--7 2022. 

Figure~\ref{fig:all_instruments}(f) shows the UV auroral power zoomed in on this event. The peak power of 12 TW was reached at 22:00 on December 6 and the full width at half maximum of the brightening event was approximately 6 hours. The maximum power output was notably high; between March 2022 and March 2024, 4200 hours (175 days) of UVS synoptic observations were obtained, and there were only six occasions when the peak auroral power exceeded 10 TW. Due to the geometrical limitations of observing Jupiter's southern aurora from Earth's orbit, there are limited datasets to directly compare this power to. However, the most powerful northern aurora ever observed by the Hubble Space Telescope, as of 2017, was 5.5 TW~\cite{nichols17}, while the highest northern auroral power reported by the Hisaki mission was 10 TW~\cite{kimura18}. 

Based on the in situ crossing measurements, we know that the compression event was already underway by 13:25 on December 6, and the UV power time series shows that the auroral brightness started to gradually increase from 6:00 onward. The peak observed auroral brightness at 22:00 likely occurred within a few hours of the maximum compression; as noted in Section~\ref{sec:size}, the maximum recorded state of compression was at 17:20 when the spacecraft crossed back into the magnetosheath from the solar wind, but actual peak compression could have occurred shortly before or in the hours after this crossing. 

The bKOM radio emission produced by Jupiter's auroras also shows a clear increase during this compression event, as shown in Figure~\ref{fig:all_instruments}(e). This correlation between enhanced bKOM radio emissions and compressions of Jupiter's magnetosphere has previously been observed in Juno data by~\citeA{louis23}. We do note that unlike the UV emissions, the peak intensity observed in the bKOM intensity is not unusually bright; while 2 nWm\textsuperscript{-2}Hz\textsuperscript{-1} is significantly higher than the baseline level,~\citeA{louis23} shows that the bKOM intensity regularly reaches this value during compression events. The radio emissions reach a peak at 20:00, shortly after the second bow shock crossing. This peak coincides with a temporal gap in the UVS coverage, as shown in Figure~\ref{fig:all_instruments}(f); although the peak observed UV emission occurred two hours later at 22:00, it is possible that the true maximum coincided with the radio emission peak.


\section{Discussion}
\label{sec:discussion}

On December 6--7 2022, the following events occurred:

\begin{enumerate}
    \item An interplanetary shock was predicted to arrive at Jupiter, causing an abrupt increase in the solar wind dynamic pressure to a level not previously experienced during the Juno mission. This increase in dynamic pressure lasted $\sim$4 hours.
    \item Jupiter's magnetosphere and magnetosheath experienced an extreme compression event, with a compression strength well above the 90th percentile, based on models from~\citeA{joy02}. The compression and expansion phases of this event together lasted over 24 hours. The magnetosphere was in the process of compressing at 13:25 on December 6, the peak compression of the dusk flank of Jupiter's magnetosphere occurred at approximately 16:00--18:00 on December 6, and the magnetosphere was still expanding at 13:15 on December 7. 
    \item The H\textsubscript{2} UV emission from Jupiter's southern aurora reached 12 TW, a factor of six higher than the baseline level. This spike in UV auroral emission lasted $\sim$6 hours. The bKOM radio emission produced by Jupiter's auroras also peaked at this time. The peak auroral response occurred at 20:00--22:00 on December 6, approximately 4 hours after the peak compression measured at the location of the Juno spacecraft.
\end{enumerate}

The powerful interplanetary shock, the strong compression event and the bright UV aurora are each individually rare events, so their simultaneous occurrence points to a clear relationship between them. Our observations are therefore consistent with an externally-forced mechanism for these UV auroral emissions, where a large solar wind shock causes the magnetosphere to dramatically compress, which in turn causes a large spike in the auroral power. This is consistent with results from previous monitoring campaigns of Jupiter's auroras, which have shown a correlation between auroral activity and solar wind conditions in both the UV~\cite{gurnett02,nichols07,clarke09,kita16,nichols17,yao22} and at other wavelengths~\cite{gurnett02,wibisono20,louis23,sinclair23,odonoghue25}.

Theoretical studies have suggested that there should be an anticorrelation between solar wind dynamic pressure and UV auroral brightness~\cite{southwood01,cowley03}. This is due to the expected increase in plasma angular velocities under a compressed state, which in turn is expected to decrease the field-aligned current (FAC) intensity and therefore decrease the UV auroral emissions. This appears to contrast with the results of both this study and the previous papers that have shown a correlation between the magnetosphere being compressed and enhanced UV auroral brightness. However, \citeA{nichols07} have pointed out that there are typically timing uncertainties, which can confuse the situation; it is difficult to tell whether it is the compression that is causing the auroral enhancement, or whether the true cause is the expansion that takes place as the magnetosphere returns to its pre-compression state.

It is clear that better understanding of the timing of these events is key to understanding the physical processes, and one advantage of the multi-instrument dataset that we have presented in this paper is that we have more detailed timing information than is often available: quasi-continuous coverage of the UV auroral emissions and four in situ measurements of the instantaneous compression state of the magnetosphere (two magnetopause crossings and two bow shock crossings). Our results show that there is a $\sim$4 hour delay between the peak magnetospheric compression measured on the southern dusk flank and the peak auroral response. There are three plausible scenarios for the causal relationship between the solar-wind-driven compression event and the auroral response:

\begin{enumerate}[(i)]
    \item The initial expansion of the magnetosphere after the peak compression causes UV auroral brightening.
    \item The compression of the nose and/or flank of the magnetosphere causes UV auroral brightening, but the auroral response is delayed by several hours.
    \item The compression of the magnetospheric tail causes near-instantaneous UV auroral brightening.
\end{enumerate}

At the location of the Juno spacecraft, our measurements of the magnetopause and bow shock crossings show that the magnetosphere was likely beginning to expand again at the time of the peak UV auroral emissions. This could support scenario (i), and this scenario would be consistent with the models of ~\citeA{southwood01} and~\citeA{cowley03}. However, it is not clear why the enhanced auroral emissions would be so restricted to the very beginning of the expansion phase; the magnetosphere was still expanding at 13:15 on December 7 when Juno crossed the magnetopause for the second time, but by this time, the UV auroral power had returned to its baseline levels. 

Instead, in scenarios (ii) and (iii), it is the compression of the magnetosphere that causes the enhancement in the auroral emissions. In scenario (ii), the compression of the nose and/or flank drive the emissions, but there is a lag time of several hours before the aurora responds. In scenario (iii), it is the compression of the magnetospheric tail that causes a more immediate response; assuming a radial solar wind speed of 470 km/s (as predicted by SWMF-OH), the $\sim$4-hour delay between the peak compression measured at Juno and the peak auroral response equates to a tailward distance of $\sim$100 R\textsubscript{J} from Jupiter.

Although both of these scenarios run counter to the expectations from the classical steady-state theoretical predictions~\cite{southwood01,cowley03}, there have been several more recent suggestions for the mechanisms by which this could occur. \citeA{cowley07} suggested that very large, rapid-onset compression events, like the compression event described in this paper, could cause super-rotation to be induced on closed field lines, resulting in a reversal of the usual magnetosphere-ionosphere coupling current system and an increase in the precipitating electron power. Alternatively,~\citeA{feng22} used a global MHD model to show that solar wind compressions lead to increases in both FAC and Alfvénic power (AP), which could explain the observed auroral responses without requiring a current reversal. The~\citeA{feng22} model produces an up-down-up response in the FAC and AP.
The initial increase in FAC and AP is relatively sudden and is due to the significant change in shape of the magnetosphere. It occurs a few hours after the solar wind compression onset, with the delay depending on the magnetosphere size and the Alfvénic transit time. The second increase, which occurs a day or two after the compression onset and is more gradual, is due to enhanced magnetospheric convection and enhanced magnetotail activity. As the auroral enhancement we observe is both short-lived and occurs a few hours after compression onset, it is more consistent with scenario (ii) and the first "up" in the up-down-up response. 

\section{Conclusions}

In this paper, we presented the response of Jupiter’s ultraviolet aurora to an extreme magnetospheric compression on December 6--7 2022. During this event, the compression strength was above the 90th percentile and the ultraviolet auroral emissions reached a peak power of 12 TW, a factor of six higher than the baseline level. The compression also coincided with the predicted arrival of a powerful interplanetary shock, which was expected to cause the largest increase in the solar wind dynamic pressure seen thus far during the Juno mission. The simultaneous occurrence of the interplanetary shock, the extreme compression and the bright ultraviolet aurora suggests that in this case, the auroral brightening was externally-driven and was caused by the solar wind shock. It is still not clear whether it is the period of magnetospheric compression itself that leads to the enhanced emissions, or whether it is the period of magnetospheric expansion that takes place immediately after the compression. However, the fact that the auroral power returns to baseline levels while the magnetosphere is still in the middle of its expansion phase leads us to favor the former, and the observed timing is consistent with theoretical predictions from global MHD models that suggest a compression-driven auroral enhancement mechanism~\cite{feng22}.  

While this paper presents a case study of a very strong UV auroral brightening event that was likely caused by a powerful solar wind shock compressing the magnetosphere, this does not mean that all auroral brightening events are triggered by the same mechanism, or that all compression events lead to increased UV auroral emissions. Previous studies have shown that UV auroral brightness enhancements do not always correlate with periods of increased solar wind~\cite{clarke09,kimura15,gladstone17b}. UVS synoptic observations of Jupiter's southern aurora have been obtained regularly since 2022, and the orbital geometry of the Juno mission means that these observations frequently coincide with detections of magnetopause crossings. Future work will compare the full time series of these two datasets to search for a statistical relationship between the total UV emissions and the compression state of the magnetosphere.


\section*{Acknowledgements}

We are grateful to NASA and contributing institutions, which have made the Juno mission possible. This work was funded by NASA's New Frontiers Program for Juno via contract with the Southwest Research Institute. CKL acknowledges support from CNES and CNRS/INSU national programs of planetology (PNP) and heliophysics (PNST).


\section*{Data Availability Statement}

The Juno data used in this paper are archived in the NASA Planetary Data System. The Juno UVS data used in this paper are archived in \citeA{trantham14}. The Juno Waves data used in this paper are archived in \citeA{kurth24}. The Juno JADE data used in this paper are archived in \citeA{allegrini22}. The data used to produce the figures in this paper are available in \citeA{giles25_data}.



\end{document}